\documentclass{Interspeech}
\interspeechcameraready 
\title{Analysis and Evaluation of Synthetic Data Generation in Speech Dysfluency Detection}

\author[affiliation={1}]{Jinming}{Zhang}
\author[affiliation={1}]{Xuanru}{Zhou}
\author[affiliation={2}]{Jiachen}{Lian}
\author[affiliation={1}]{Shuhe}{Li}
\author[affiliation={2}]{William}{Li}
\author[affiliation={3}]{Zoe}{Ezzes}
\author[affiliation={3}]{Rian}{Bogley}
\author[affiliation={3}]{Lisa}{Wauters}
\author[affiliation={3}]{Zachary}{Miller}
\author[affiliation={3}]{Jet}{Vonk}
\author[affiliation={3}]{Brittany}{Morin}
\author[affiliation={3}]{Maria}{Gorno-Tempini}
\author[affiliation={2}]{Gopala}{Anumanchipalli}

\affiliation{}{Zhejiang University}{China}
\affiliation{}{UC Berkeley}{USA}
\affiliation{}{UCSF}{USA}
\email{pmhuan1212@gmail.com, jiachenlian@berkeley.edu, gopala@berkeley.edu}

\usepackage{comment}
\usepackage{stfloats} 
\usepackage{placeins}
\usepackage{float} 
\usepackage{subcaption}
\usepackage{multirow}
\usepackage{caption}
\usepackage{float} 
\usepackage{adjustbox}  
\usepackage{afterpage}

\usepackage{xcolor}
\usepackage{booktabs}
\usepackage{array}
\usepackage{subcaption}
\usepackage{multirow}

\begin{document}

\maketitle
\begin{abstract}
Speech dysfluency detection is crucial for clinical diagnosis and language assessment, but existing methods are limited by the scarcity of high-quality annotated data. Although recent advances in TTS model have enabled synthetic dysfluency generation, existing synthetic datasets suffer from unnatural prosody and limited contextual diversity. To address these limitations, we propose LLM-Dys --- the most comprehensive dysfluent speech corpus with LLM-enhanced dysfluency simulation. This dataset captures 11 dysfluency categories spanning both word and phoneme levels. Building upon this resource, we improve an end-to-end dysfluency detection framework. Experimental validation demonstrates state-of-the-art performance. All data, models, and code are open-sourced at \url{https://github.com/Berkeley-Speech-Group/LLM-Dys}.

\end{abstract}
\vspace{-5pt}
\section{Introduction}
Speech dysfluency detection is an essential step for assisting in disordered speech diagnosis, language screening, or early prevention. For a long time, dysfluency or stutter detection has been treated simply as a classification problem, mostly binary~\cite{abubakar2024stutternet, shih2024self-ssl-stutter, barrett2022systematic-stutter1.0, jouaiti2022dysfluency-stutter1.1, bayerl2022detecting-stutter1.2, zayats2016disfluency-stutter1.3, alharbi2017segment-detection2, alharbi2020segment-detection3, harvill2022frame-level-stutter, shonibare2022frame-detection2, wagner2024large}, among others. However, to better serve clinical needs, researchers have developed more sophisticated methods~\cite{UDM, lian-anumanchipalli-2024-towards, zhou24e_interspeech} that can identify both the types and timing of dysfluencies.

The development of robust dysfluency detectors requires large-scale, high-quality datasets. While public corpora like UCLASS~\cite{howell2009university} and SEP-28K~\cite{lea2021sep} exist, they have limitations in both size and annotation quality. The segmentation and annotation in these datasets often fall short of the requirements for training robust models. For instance, SEP-28K contains numerous partially pronounced words and lacks accurate ground truth transcriptions, making it challenging to develop reliable dysfluency detection systems. Some researchers have attempted to simulate dysfluent speech. For example, LibriStutter~\cite{kourkounakis2021fluentnet}, VCTK++~\cite{UDM}, and~\cite{harvill2022frame-level-stutter} directly inject dysfluencies in the time or spectrogram domain. However, this approach has been shown to produce low audio quality~\cite{zhou24e_interspeech}. Instead, \cite{zhou24e_interspeech} proposed VCTK-TTS, where dysfluencies are simulated only at the text level and then synthesized into speech using a TTS model~\cite{kim2021conditional}. VCTK-TTS demonstrates significantly better intelligibility and naturalness than all previous datasets. A subsequent study~\cite{zhou2024stutter-solver} introduced VCTK-Pro by incorporating co-dysfluencies. Libri-Dys~\cite{ssdm} adopted the same technology but further scaled it up to LibriTTS~\cite{zen2019libritts}, and then extended it to co-dysfluency~\cite{lian2024ssdm2.0}. Additionally, VCTK-Token~\cite{zhou2024timetokensbenchmarkingendtoend}shares the same simulation pipeline as VCTK-TTS but includes token-level labels.
A key shortcoming of these TTS-based simulated corpora is that the text simulation is purely rule-based, which may \textit{not accurately reflect human stuttering patterns}. Moreover, \textit{the diversity of text variations}~\cite{yamagishi2019cstr, zen2019libritts} \textit{explored in these methods remains quite limited}. Another issue is that there is no standard for dysfluent speech labeling. For traditional binary classification-based detection, the label is simply "stutter" or "not stutter". For advanced clinical-aware dysfluency modeling~\cite{UDM, zhou24e_interspeech}, both types and accurate timestamps for all dysfluency types are usually required. For normal ASR tasks, the dysfluency labels can just be limited to filler words~\cite{wagner2024crisperwhisper}. \textit{The lack of a high-quality, text-diversified, naturalistic corpus with unified labels} makes scaling efforts particularly challenging. 

In this work, we propose leveraging Large Language Models (LLMs) to generate dysfluent text across a diverse textual corpus, capitalizing on their learned understanding of natural dysfluency patterns. The generated text is synthesized using a Text-to-Speech (TTS) model to create \textit{LLM-Dys}, which constitutes the largest simulated dysfluency corpus to date, with over \textit{10,000 hours of speech} (as presented in Table \ref{table_1}). Recognizing that traditional binary stuttering detection can be viewed as a subset of the broader multi-class dysfluency localization problem, we focus on modeling eleven distinct dysfluency types: insertions, repetitions, pauses, deletions, and substitutions at both word and phoneme levels, as well as phonetic prolongations. This is the \textit{first comprehensive and opensourced dataset covering all major types of dysfluencies} and can be adapted to a wide range of dysfluency detection tasks. 

Quantitative analysis~\cite{tjandra2025meta} reveals that LLM-Dys achieves superior synthesis quality compared to other text-diversity-constrained simulation corpora and is even comparable to real fluent speech, as visualized in Fig~\ref{fig:eval1}. We perform dysfluency detection on both our simulated and real stuttered speech benchmarks, consistently achieving state-of-the-art performance. To further explore the limits of text-based simulation, we investigate scaling laws~\cite{kaplan2020scaling} with respect to data and report that simply increasing textual data or diversity may not yield additional performance improvements unless a high-quality acoustic (TTS) model is employed. Additional ablation studies demonstrate that our proposed benchmark is generalizable and robust, and we hope it will further facilitate research in the community.

\vspace{-7pt}
\section{Data Simulation}  
\label{sec:data simulation}
\subsection{Dysfluent Text Generation}
\label{subsec:dysfluent-generation}
We build upon the most recent and naturalistic simulated corpus VCTK-Token~\cite{zhou2024timetokensbenchmarkingendtoend}, leveraging Large Language Models instead of rule-based methods to generate authentic dysfluent texts. Through prompting, we obtain both dysfluent texts and corresponding labels, eliminating manual annotation needs (The label can be adjusted according to specific tasks). Our LLM implementations is based on \texttt{claude-3-5-sonnet}~\cite{claude3}. The prompts we used can be found at our open-sourced page. 
When generate phoneme-level utterances, we also provide clean texts with their CMU and IPA sequences (via phonimizer~\cite{Bernard2021}) as additional context, enabling LLMs to generate phonetically valid dysfluent sequences.

\begin{figure*}[t]
    \begin{adjustbox}{center}
    \includegraphics[width=15.0cm]{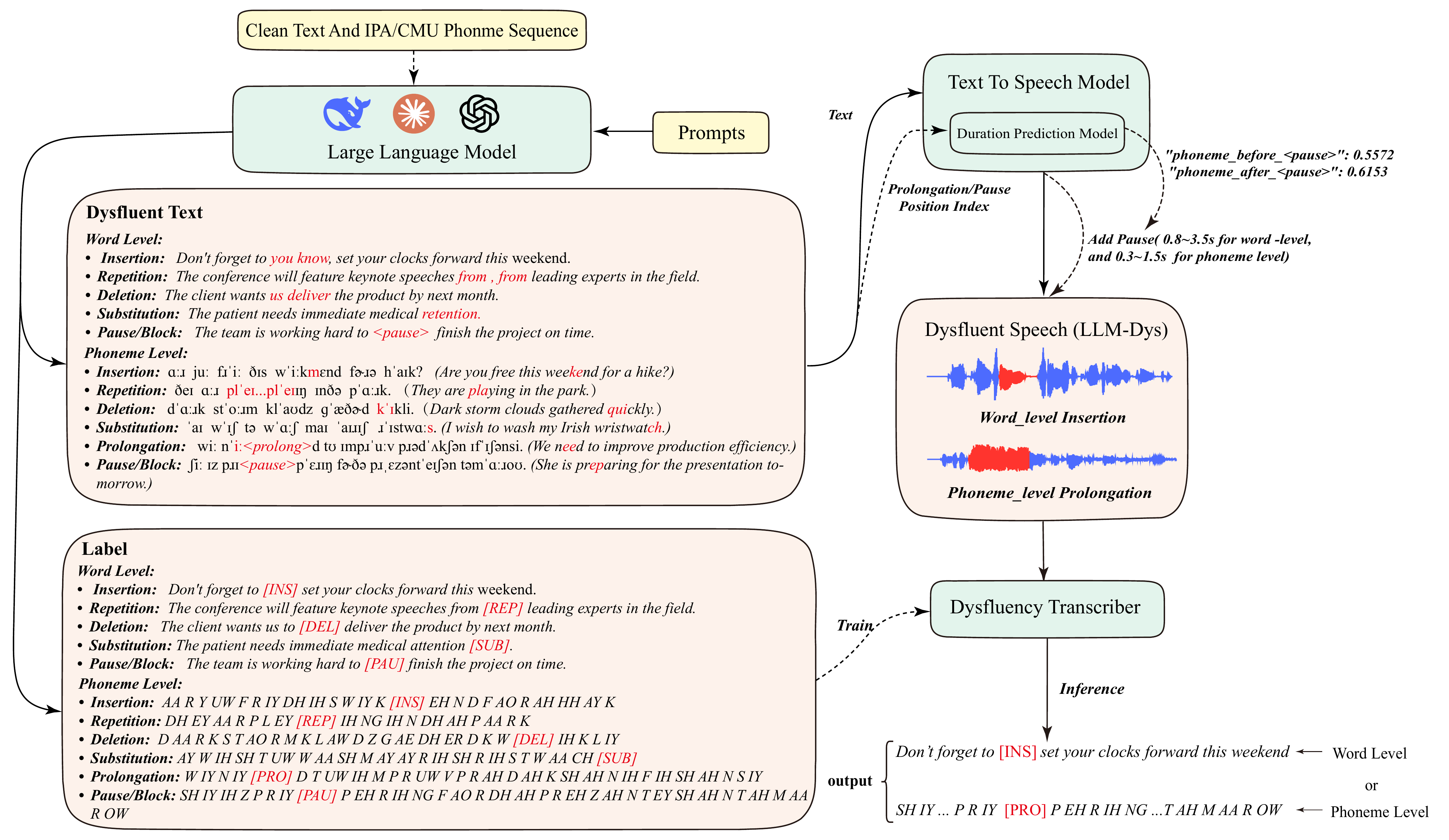}
    \end{adjustbox}
    \caption{Overview of our approach: We leverage Large Language Models (LLMs) to generate dysfluent text for TTS synthesis and corresponding labels for dysfluency transcriber training. After applying special processing for pauses and prolongations, we establish a large-scale dataset called LLM-Dys. By jointly feeding acoustic features and labels into the dysfluency transcriber for training, we achieve end-to-end dysfluent speech detection. }
    \vspace*{-0.5cm}   
    \label{figure_overview}
\end{figure*}

\begin{table*}[b]

\caption{Utterance and Duration statistics of our synthetic dataset LLM-Dys. In this table, '10028*109' means there're 10028 different utterances generated by LLMs, and 109 speakers are used for Synthesis}

\label{table_1}
\centering
\setlength{\tabcolsep}{5pt}
\renewcommand{\arraystretch}{1.0}
\begin{tabular}{c c c c c c c c}
\toprule
Level & Types & Insertion & Repetition & Pause & Deletion & Substitution & Prolongation \\
\hline
\hline
\multirow{2}{*}{\centering Word} & Samples & 10028*109 & 14184*109 & 7667*109 & 10000*109 & 9876*109 & - \\
{\it \rule{0pt}{1pt}(109 speakers)} & Hours & 1540 hrs & 1916 hrs & 1379 hrs & 1140 hrs & 868 hrs & - \\
\hline
\hline
\multirow{2}{*}{\centering Phoneme} & Samples & 9298*109 & 9377*109 & 9396*109 & 8917*109 & 6858*109 & 9500*109 \\
{\it \rule{0pt}{7pt}(109 speakers)} & Hours & 1008 hrs & 1021 hrs & 1742 hrs & 732 hrs & 499 hrs & 945 hrs \\
\bottomrule

\end{tabular}
\end{table*}

\vspace{-5pt}
\subsection{Dysfluent Speech Generation}
\label{subsec:speech-synthesis}
\vspace{-3pt}
We primarily adopt VITS~\cite{kim2021conditional} for dysfluent speech generation. Our experiments show that VITS is more reliable in generating dysfluent speech, particularly in preserving dysfluent segments rather than automatically omitting them.  Furthermore, with specific modifications, VITS can directly accept IPA sequences as input, enabling phoneme-level dysfluency simulation. Its robust duration prediction model allows precise timestamp insertion for pauses and accurate prolongation of specified phonemes. However, VITS shows limitations in synthesizing fillers like "um" and "uh," which constitute a significant portion of inserted filler words. Therefore, we employ E2-TTS~\cite{chen-etal-2024-f5tts,10832320} for word-level insertions. VITS includes 109 VCTK speakers, generating 109 samples per LLM-generated utterance. For E2-TTS, which requires reference audios, we extract sample clips from each VCTK speaker. This allows us to generate an equivalent set of 109 variations per LLM-generated utterance, ensuring dataset consistency. Some examples of LLM-generated utterances are shown in Fig~\ref{figure_overview}.
We provide explanations for pause and prolongation implementations:
\vspace{-2pt}

\begin{enumerate}[label=\textbullet]
    \item \textbf{Pause:} 
    LLMs first generate \texttt{<pause>} markers in the dysfluent text. We then generate fluent speech and obtain timestamps for the phonemes adjacent to the \texttt{<pause>} marker. We then smoothly insert a silent segment of 0.8-3.5s (word level) or 0.3-1.5s (phoneme level) into the fluent speech. 
    \item \textbf{Prolongation:} Using the \texttt{<prolong>} markers generated by LLMs, we identify the prolong position index, which corresponds to the position of the target prolonged phoneme in the VITS duration matrix. During audio synthesis with VITS, we extend the duration of this phoneme by 0.17-0.8s.

\end{enumerate}

\begin{figure*}[!t]  
    \vspace*{-0cm}   
    \begin{minipage}{\textwidth}
        \centering
        \includegraphics[width=0.85\textwidth]{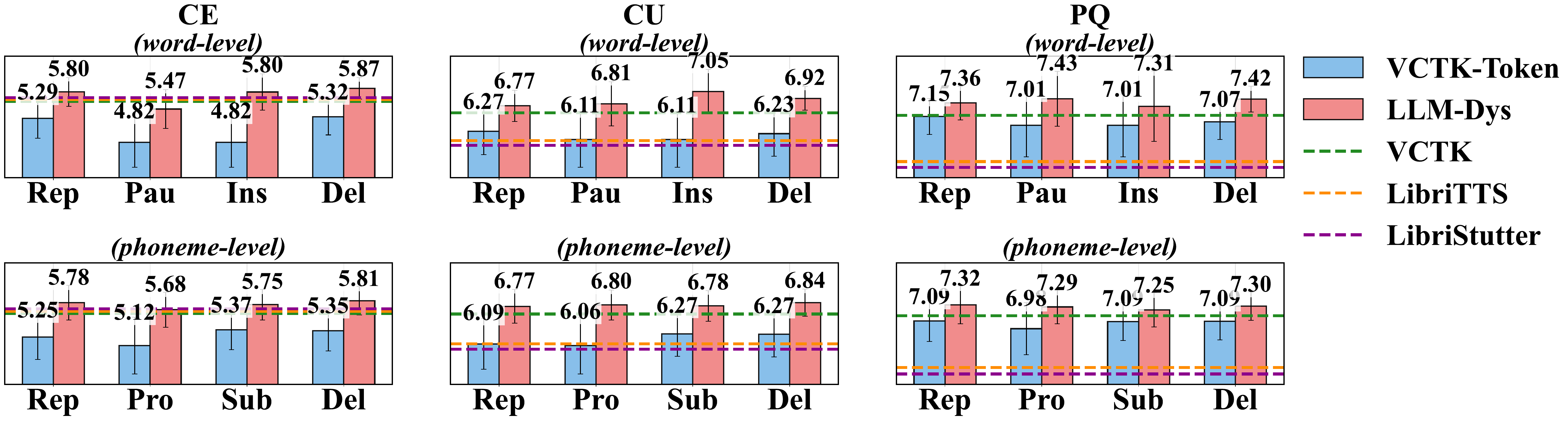}
        \caption{Comparision between differenct datasets, CE:Content Enjoyment, CU:Content Usefulness, PQ: Production Quality}
    \label{fig:eval1}
    \end{minipage}

    \vspace{0.4cm} 

    \begin{minipage}{\textwidth}
        \centering
        \includegraphics[width=0.85\textwidth]{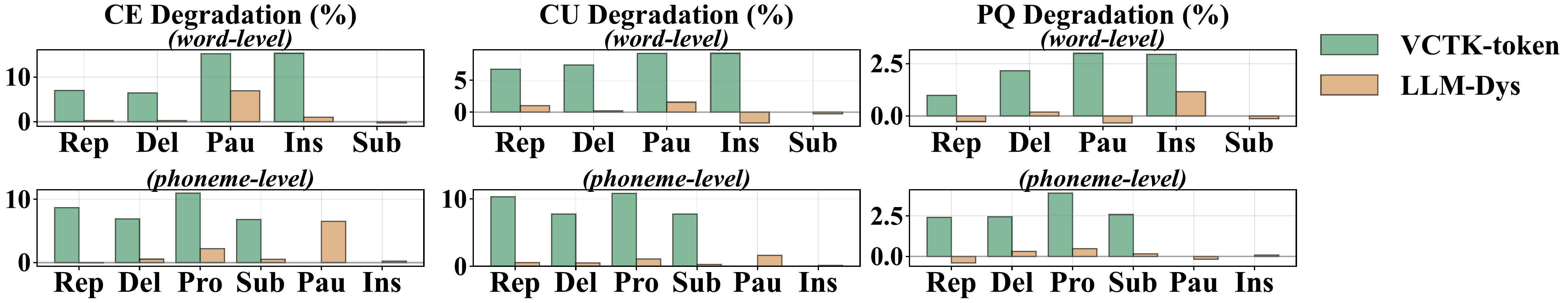}
        \caption{Comparative quality degradation analysis between LLM-Dys and VCTK-token}
        \vspace*{-0.5cm}   
        \label{fig:combined_performance_comparison}
        
    \end{minipage}
\end{figure*}

\begin{figure}[ht]
    \begin{adjustbox}{center}
    \includegraphics[width=0.25\textwidth]{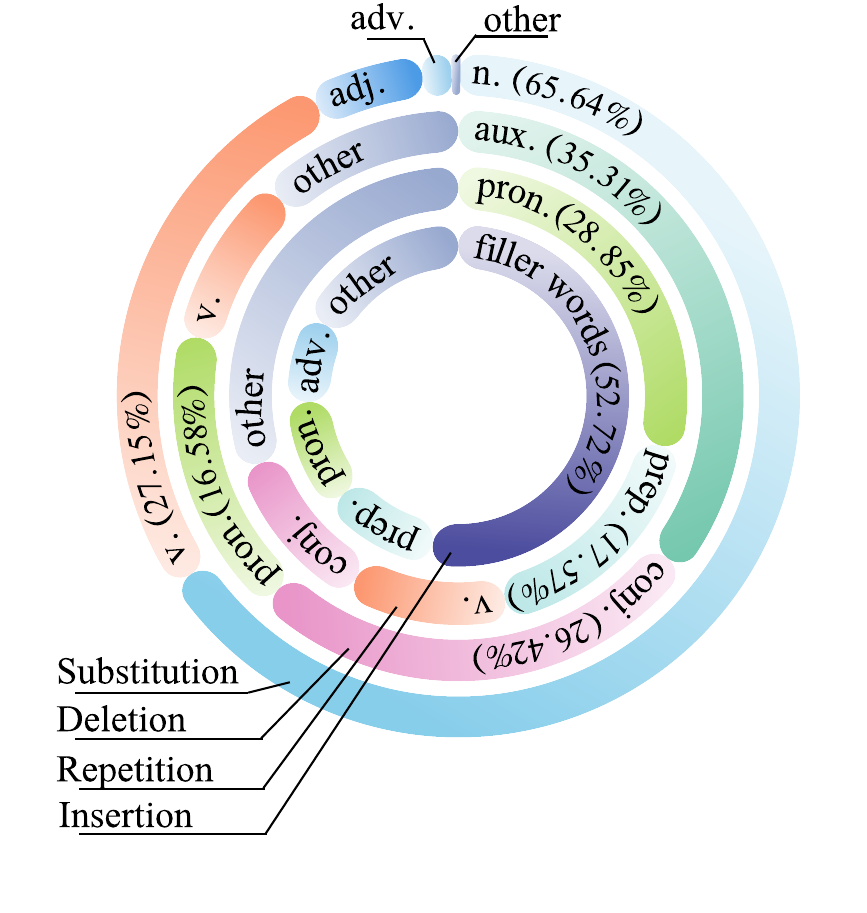}
    \end{adjustbox}
    \vspace{-10pt}
    \caption{POS analysis on four word level subsets of LLM-Dys}
    \vspace{-10pt}
    \label{figdata_cate}
\end{figure}

\vspace{-8pt}
\subsection{Statistics}
\label{subsec:statistucs}
\vspace{-3pt}
As detailed in Table~\ref{table_1}, we generate utterances per type using LLMs.
The total duration of word-level and phoneme-level speech amounts to approximately 6,843 and 5,947 hours respectively, resulting in a substantial dataset of \textbf{12,790 hours}. 


In Fig~\ref{figdata_cate}, We analyze POS patterns for four word-level dysfluency types. The analysis reveals distinct LLM-generated patterns: (1) Substitutions: LLMs exchange words with similar pronunciations, particularly nouns and verbs; (2) Deletions: commonly occur with auxiliary verbs and conjunctions,which mirrors natural speech patterns where speakers often omit these elements for efficiency; (3) Repetitions: LLMs frequently duplicate pronouns and prepositions, reflecting natural hesitation points; (4) Insertions: primarily use filler words, simulating natural speech pauses and thinking time.

\vspace{-9pt}
\subsection{Dataset Evaluation}
\label{subsec:dataset-evaluation}
\vspace{-4pt}

To comprehensively and objectively evaluate our dataset, particularly the naturalness of speech, we employ \textit{Meta Audiobox Aesthetics} ~\cite{tjandra2025meta} as our evaluation tool. This tool can directly assess input audio samples across multiple dimensions and provides four metrics: Content Enjoyment (CE), Content Usefulness (CU), Production Complexity (PC), and Production Quality (PQ). Among these metrics, we specifically selected CE, CU, and PQ as they effectively reflect the overall quality and naturalness of our dataset. (We exclude PC from our analysis as it primarily measures the number of audio components, which is less relevant for our test samples where each audio clip contains only single-speaker utterances.)

\vspace{-9pt}
\subsubsection{Cross-dataset Comparison of Absolute Audio Quality}
\vspace{-3pt}

We evaluate 2,000 random samples from each comparable category across LLM-Dys, VCTK-token, VCTK, LibriTTS, and LibriStutter~\cite{kourkounakis2021fluentnet}. Results show LLM-Dys achieves superior performance across almost all metrics compared to both fluent (VCTK, LibriTTS) and dysfluent speech datasets, as shown in Fig~\ref{fig:eval1}.

\vspace{-5pt}
\subsubsection{Analysis of Speech Quality Before and After Dysfluency}
\vspace{-3pt}
To further validate our methodology, we conduct a comparative analysis of speech quality metrics. We synthesize both the clean text and its corresponding dysfluent version using TTS, then compare their metrics to calculate the degradation rates. As shown in Fig \ref{fig:combined_performance_comparison}, LLM-Dys achieves better metric preservation and even slight improvements in certain categories, demonstrating its superior performance over rule-based approaches in maintaining speech naturalness and quality while introducing dysfluencies.

\vspace{-6pt}
\section{Token-based Dysfluency Detection}  

\label{sec:Token-based Dysfluency Detection}
\vspace{-3pt}
We follow \cite{zhou2024timetokensbenchmarkingendtoend} to treat dysfluency detection as a token-based recognition problem and adopt Whisper-large-v3-turbo\cite{radford2023robust} as our base model.  We divide dysfluency detection into word and phoneme levels. Based on the annotated dysfluency types in SEP-28K dataset, at word\_level, we train for insertion, pause, and repetition using  word\_ins, word\_pau, and word\_rep subsets from LLM-Dys with a 1:1:1 ratio. During training, we incorporate a proportion of VCTK dataset, which is explained in Section \ref{scaling}.
At phoneme\_level, we train for prolongation, pause, and repetition using phn\_pro, phn\_pau, and phn\_rep subsets from LLM-Dys with a 1:1:1 ratio. Notably, we observe that SEP-28K dataset contains relatively few samples of phoneme-level pause dysfluency (referring to pauses occurring within words, such as "dys...dysfluency"). Therefore, during training and testing, we supplement phoneme-level pause samples with some word-level pause samples (phoneme:word = 3:7, all annotations are based on phoneme-level) to balance different types of dysfluency. Additionally, we incorporate a proportion of VCTK dataset as well.

\begin{table}[t]
 \caption{Metrics on LLM-Dys}
  \vspace{-5pt}
 \label{eeval_llmval_llm}
\centering
\resizebox{\textwidth/2}{!}{
\begin{tabular}{c|c|ccc|ccc}
\toprule
\multirow{2}{*}{Model} & \multirow{2}{*}{Metrics} & \multicolumn{3}{c|}{Word Level} & \multicolumn{3}{c}{Phoneme Level} \\
\cmidrule(lr){3-5} \cmidrule(lr){6-8}
& & Ins & Rep & Pau & Pau & Rep & Pro \\
\midrule
\multirow{5}{*}{\shortstack{Ours \\ (3*4000 samples)}} 
& Recall & 0.99& 0.99& 1.0& 0.99& 1.0& 0.99\\
& Precision & 1.0& 1.0& 0.99& 1.0& 1.0& 1.0\\
& F1-score & 0.99& 0.99& 1.0& 0.99& 1.0& 0.99\\
& TER($\%$, $\downarrow$) & 4.63& 2.52& 2.54& 0.78& 1.04& 0.72\\
& TD($\downarrow$)& 0.76& 0.22& 0.52& 0.18& 0.33& 0.10\\
\bottomrule
\end{tabular}
}
\end{table}

\begin{table}[!t]
 \caption{Precision, Recall and F1-score on SEP-28k}
 \vspace{-5pt}
 \label{eval_sep_1}
 \renewcommand{\arraystretch}{0.7}  
\setlength{\tabcolsep}{5pt}      
\centering
\resizebox{\textwidth/2}{!}{
\begin{tabular}{c|c|ccc|ccc}
\toprule
\multirow{2}{*}{Model} & \multirow{2}{*}{Metrics} & \multicolumn{3}{c|}{Word Level} & \multicolumn{3}{c}{Phoneme Level} \\
\cmidrule(lr){3-5} \cmidrule(lr){6-8}
& & Ins & Rep & Pau & Pau & Rep & Pro \\
\midrule
\multirow{3}{*}{\shortstack{Ours \\ (3*4000 samples)}} 
& Recall & 0.87& 0.91& 0.71& 0.75& 0.90& 0.85\\
& Precision & 0.95& 0.52& 0.89& 0.75& 0.92& 0.97\\
& F1-score  & \textbf{0.91}& 0.67& 0.79& \textbf{0.75}& 0.91& 0.69\\

\midrule
\multirow{3}{*}{\shortstack{Ours \\ (3*12000 samples)}} 
& Recall & 0.91& 0.79& 0.71& 0.71& 0.90& 0.8
\\
& Precision & 0.86& 0.59& 1.00& 0.74& 0.96& 0.97\\
& F1-score  & 0.89& \textbf{0.68}& \textbf{0.83}& 0.72& \textbf{0.93}& \textbf{0.88}\\
\midrule
\midrule
Wagner et al.~\cite{wagner2024large} & F1-score& 0.77 & 0.64 & 0.62 & 0.62 & 0.54 & 0.56 \\
\midrule
\midrule
Yolo-Stutter~\cite{zhou24e_interspeech} & Recall & - & 0.82 & 0.72& 0.72 & - & 0.89\\
\bottomrule
\end{tabular}
}
\end{table}

\begin{table}[!t]
 \caption{Token Error Rate and Token Distance on SEP-28k}
  \vspace{-5pt}
 \label{eval_sep_2}
\centering
\renewcommand{\arraystretch}{0.8}  
\setlength{\tabcolsep}{6pt}      
\resizebox{\textwidth/2}{!}{
\begin{tabular}{c|c|ccc|ccc}
\toprule
\multirow{2}{*}{Model} & \multirow{2}{*}{Metrics} & \multicolumn{3}{c|}{Word Level} & \multicolumn{3}{c}{Phoneme Level} \\
\cmidrule(lr){3-5} \cmidrule(lr){6-8}
& & Ins & Rep & Pau & Pau & Rep & Pro \\
\midrule
\multirow{2}{*}{\shortstack{Ours \\ (3*4000 samples)}} 
& TER($\%$, $\downarrow$) & 24.90& 22.32& 16.27& \textbf{7.12}& \textbf{11.68}& \textbf{11.05}\\
& TD( $\downarrow$)& 1.17& \textbf{0.27}& 1.59& \textbf{1.06}&\textbf{ 1.50}& 1.38\\

\midrule
\multirow{2}{*}{\shortstack{Ours \\ (3*12000 samples)}} 
& TER($\%$, $\downarrow$) & \textbf{23.10}& \textbf{19.8}4& \textbf{15.88}& 9.89& 12.17& 11.55\\
& TD($\downarrow$)&\textbf{ 0.75}& 0.47&\textbf{ 1.24}& 1.76& \textbf{1.50}& \textbf{1.28}\\
\midrule
\bottomrule
\end{tabular}
}
\end{table}
\begin{table}[!t]
\caption{Accuracy, Precision, Recall, and F1-score on UCLASS}
\label{tab:results_ucalss}
\centering
\renewcommand{\arraystretch}{0.6}  
\setlength{\tabcolsep}{5pt}      
\resizebox{\textwidth/2}{!}{
\begin{tabular}{c|c||c|c|c|c}
\toprule
Model & Level & Accuracy & Precision & Recall & F1-score \\
\midrule
\multirow{2}{*}{\shortstack{Ours \\ (3*12000 samples)}} 
& Word & \textbf{0.971}& \textbf{1.000}& 0.954& \textbf{0.977}\\
& Phoneme & 0.958& 0.938& \textbf{1.000}& 0.968\\
\midrule
StutterNet~\cite{abubakar2024stutternet} & - & 0.938& 0.931& 0.933& 0.932\\
\bottomrule
\end{tabular}
}
\end{table}

\vspace{-5pt}
\section{Experiments} \label{sec:experiments}
\vspace{-5pt}
\subsection{Datasets}
\vspace{-5pt}

\textbf{1) LLM-Dys}: Our synthetic dataset contains 11 dysfluency types, totaling 12,790 hours. Details in Section~\ref{sec:data simulation}.\textbf{2) SEP-28k}~\cite{lea2021sep}: Real-world dataset with 28,000 clips, labeled with blocks, prolongations, sound/word repetitions, and interjections. Due to poor segmentation quality, we created a test set by manually annotating 200 samples each for word/phoneme-level evaluation, maintaining the same distribution of dysfluency types as in the original dataset. Annotations follow our model's output format with dysfluency tokens added to clean text.
\textbf{3) UCLASS}~\cite{howell2009university}:Speech recordings from 128 stuttering children and adults. We randomly segmented 200 samples from this dataset, 80 for fine-tuning and 120 for testing, labeled binary (fluent/dysfluent) with 1:1 ratio as in \cite{abubakar2024stutternet}.
\textbf{4) VCTK}~\cite{yamagishi2019cstr}: Natural speech corpus from 110 English speakers with diverse accents.
\textbf{5) LibriTTS}~\cite{zen2019libritts}: High-quality synthetic dataset derived from LibriSpeech.

\vspace{-6pt}
\subsection{Metrics}
\vspace{-4pt}

\textbf{1) Recall}: Ratio of correctly identified to total actual disfluencies.
\textbf{2) Precision}: Ratio of correctly identified to total predicted disfluencies.
\textbf{3) F1-score}: Harmonic mean of precision and recall.
\textbf{4) Accuracy (Acc)}: Model's performance in identifying fluent speech and dysfluency types.
\textbf{5) Token Error Rate (TER)}: Transcription accuracy compared to reference text, similar to WER.
\textbf{6) Token Distance (TD)}: Token-level displacement between predicted and actual dysfluency positions.

\vspace{-6pt}
\subsection{Results}
\vspace{-3pt}

\subsubsection{Evaluation on LLM-Dys}
\vspace{-4pt}
We test on 300 unique utterances per dysfluency type from testing set. As shown in Table \ref{eeval_llmval_llm}, the model achieves high performance  despite limited training data (4,000 samples/type), likely due to consistent patterns in LLM-generated dysfluencies and our standardized TTS pipeline. 
\vspace{-5pt}

\subsubsection{Evaluation on SEP-28k and UCLASS}
\label{eval_pubic}
\vspace{-3pt}
For SEP-28k, we conduct zero-shot evaluation (which inherently puts our model at a disadvantage compared to~\cite{wagner2024large}) and still achieve state-of-the-art results, as shown in Table~\ref{eval_sep_1} and Table~\ref{eval_sep_2}. Since the original SEP-28k annotations only contain block labels (without distinguishing between word-level and phoneme-level), we apply the block-level scores reported in~\cite{wagner2024large} to both word-level and phoneme-level metrics. 
For UCLASS, we freeze the LLM-Dys fine-tuned Whisper encoder and add a classification head with three FC layers (512→256→2) for binary fluency detection. Fine-tuned with balanced samples~\cite{abubakar2024stutternet}, our model achieves SOTA performance using only 80 training clips, as shown in Table~\ref{tab:results_ucalss}.

\begin{figure}[h] 
    \begin{adjustbox}{center}
    \includegraphics[width=0.43\textwidth]{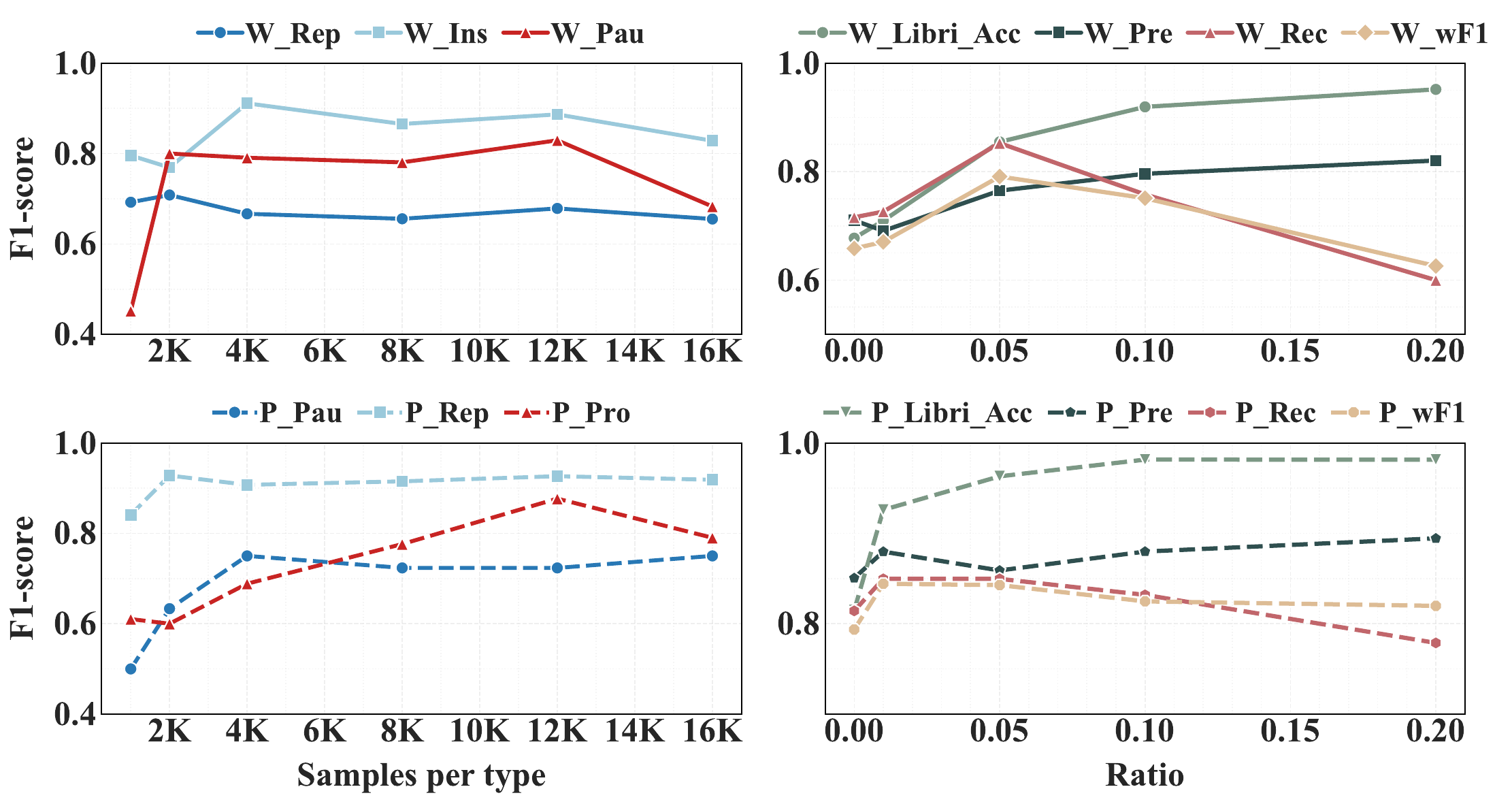}
    \end{adjustbox}
    \caption{\textbf{Left:} Impact of dataset size on dysfluency detection performance. 
    \textbf{Right:} Impact of Fluent-to-Disfluent speech ratio on model performance (P = Phoneme-level, W = Word-level, Libri = LibriTTS, Acc = Accuracy, Pre = Precision, Rec = Recall, wF1 = weighted F1 score computed based on disfluency type frequencies).}
    \label{fig:last}
\end{figure}

\vspace{-5pt}
\subsubsection{Scaling Law}
\label{scaling}
\vspace{-4pt}
Our scaling experiments reveal that the model's performance, as measured by F1 scores, reaches a substantial level with a dataset size of 3×4000 samples. Further expansion to 3×12000 samples yields only marginal improvements, after which performance plateaus or slightly declines, as illustrated in Fig.~\ref{fig:last}. These findings suggest an optimal dataset size threshold for efficient training in dysfluency detection tasks.
\vspace{-5pt}
\subsubsection{Impact of Fluent-to-Disfluent Speech Ratio}
\vspace{-4pt}
Training solely on LLM-Dys reduces fluent speech detection accuracy. Our analysis reveals that the model achieves optimal performance when the fluent-to-disfluent speech ratio is approximately 0.05 under the disfluency distribution condition of SEP-28k, as illustrated in Fig.~\ref{fig:last}

\vspace{-5pt}

\section{Conclusion and Future Work}

We introduce LLM-Dys, a large-scale dysfluency dataset spanning 11 categories and 12,790 hours. Our method generates higher-quality synthetic speech than rule-based baselines while preserving dysfluency authenticity, and achieves state-of-the-art performance on real-world dysfluency detection. Experiments reveal optimal dataset sizes and the importance of balanced fluency ratios during training. Future directions include expanding speaking styles, emotional contexts, cross-lingual coverage, and integrating articulatory priors~\cite{cho2024jstsp, wu23k_interspeech, lian22bcsnmf, lian2023factor} for improved simulation and detection.
\vspace{-6pt}
\section{Acknowledgements}
Thanks for support from UC Noyce Initiative, Society of Hellman Fellows, NIH/NIDCD, and the Schwab Innovation fund.

\vspace{-6pt}
\bibliographystyle{IEEEtran}
\bibliography{refs}

\end{document}